\documentclass[twocolumn,preprintnumbers,nofootinbib]{revtex4}
\usepackage{graphicx}
\usepackage{amssymb}
\usepackage{color}
\usepackage{enumerate}
\def\beq{\begin{equation}}
\def\eeq{\end{equation}}
\def\beqn{\begin{eqnarray}}
\def\eeqn{\end{eqnarray}}

\renewcommand{\texttt}{{}}
\newcommand{\be}{\begin{eqnarray}}
\newcommand{\ee}{\end{eqnarray}}

\begin{document}

\title{Super-renormalizable Quantum Gravity}
\author{
Leonardo Modesto}
\affiliation{Perimeter Institute for Theoretical Physics, 31 Caroline St., Waterloo, ON N2L 2Y5, Canada}
%

\date{\small\today}

\begin{abstract} \noindent
In this paper we study perturbatively an 
extension of the Stelle higher derivative 
gravity involving an infinite number of derivative terms. 
We know that the usual quadratic action is renormalizable 
but suffers of the unitarity problem because of the presence of a ghost 
(state of negative norm) 
in the theory.
The new theory is instead 
ghost-free since the introduction of (in general) 
two entire functions in the model
with the property do 
not introduce new poles in the propagator. The local high derivative theory 
is recovered expanding the entire functions to the lowest order in the mass scale of the theory. 
Any truncation of the entire functions gives rise to 
the unitarity violation but if we keep all the infinite series we do not fall into these troubles.  
The theory is renormalizable at one loop and finite from two loops on.
Since only a finite number of graphs are divergent then the theory is super-renormalizable.
We analyze the fractal properties of the theory at high energy showing a reduction of the spacetime
dimension at short scales. Black hole spherical symmetric solutions are also studied omitting the high curvature corrections in the equation of motions. The solutions are regular and the classical singularity
is replaced by a ``de Sitter-like core" in $r=0$. Black holes may show a ``multi-horizon" structure 
depending on the value of the mass.

\end{abstract}

\maketitle

\section{Introduction to the theory}

One of the biggest problems in theoretical physics is to find a theory that is able to reconcile 
general relativity and quantum mechanics. There are  many reasons to believe that 
gravity has to be quantum, some of which are: the quantum nature of matter in the right-hand side of the 
Einstein equations, the singularities appearing in classical solutions of general relativity, etc. 

The action principle for gravity we are going to introduce in this paper is the result of 
a synthesis of minimal requirements: 
(i) classical solutions must be singularity-free, 
(ii) Einstein-Hilbert action should be a good approximation of the theory at an energy scale much smaller than the Planck mass,  
(iii) the spacetime dimension has to decrease with the energy in order to have 
a complete quantum gravity theory in the ultraviolet regime, 
(iv) the theory has to be perturbatively renormalizable at quantum level 
(this hypothesis is strongly related to the previous one), 
(v) the theory has to be unitary, with no other 
degree of freedom than the graviton in the propagator, 
(vi) spacetime is a single continuum of space and time and in particular the 
Lorentz invariance is not broken. 
The last requirement is supported by recent observations.



Now let us introduce the theory, step by step, starting from the perturbative 
non-renormalizable Einstein gravity, 
through high derivatives gravity theories (the Stelle theory of gravity will be our first example)
onto the action which defines a complete quantum gravity theory.
The impatient reader can skip to the end of the introduction for the candidate complete 
quantum gravity bare Lagrangian.

Perturbative quantum gravity is the quantum theory of a spin two particle on a fixed 
(usually for simplicity is assumed to be flat) background. 
Starting from the Einstein-Hilbert Lagrangian 
\be
{\mathcal L} = - \sqrt{ - g} \, \kappa^{-2} \, R 
\ee
 ($\kappa^2 = 16 \pi G_N$) we 
introduce a splitting of the metric in a background part plus a fluctuation 
\be
\sqrt{ - g} g^{\mu \nu} =  {g}^{o \mu \nu} + \kappa h^{\mu \nu},
\ee
then we expand the action 
in power of the graviton fluctuation $h^{\mu\nu}$ around the fixed background $g^{\mu \nu}$.
Unlikely the quantum theory is divergent 
at two loops, producing a counter-term proportional to the Ricci tensor at the third power 
\be 
\sqrt{ - g} R^{\alpha \beta}_{\gamma \delta} R^{\gamma \delta}_{\rho \sigma} 
R^{\rho \sigma}_{\alpha \beta}.
\ee
In general, in $d$ dimensions the superficial degree of divergence of a Feynman diagram is 
$D = L \, d +2 V -2 I$, where $L$ is the number of loops, 
 $V$ is the number of vertices and $I$ the number of internal lines in the graph.
 Using the topological relation between $V$, $I$ and $L$, $L = 1+ I -V$, we obtain
$D = 2 + (d -2) L$. In $d=4$ the superficial degree of 
divergence $D=2+2L$ increases with the number of loops and thus we are forced to 
introduce an infinite number of higher derivative counter terms and then an infinite number of coupling 
constants, therefore making the theory not predictive.
Schematically, we can relate the loop divergences in perturbative quantum gravity to 
the counter terms we have to introduce to regularize the theory. In short 
\be
S = - \int {\rm d}^d x \sqrt{g} \, \Big[ \kappa^{-2} \, R +   \frac{\alpha}{\epsilon} 
\hspace{-0.3cm} 
\underbrace{\,\,  \sum_{m, n}  \nabla^n R^m}_{n+2 m = 2 +(d -2)L}   ,
 \hspace{-0.3cm} 
  \Big]
\ee
with ``$n$" and ``$m$" integer numbers. 

A first revolution in quantum gravity was introduced by Stelle \cite{Stelle} with the higher derivative theory 
\be
S = - \int {\rm d}^4 x \sqrt{g} \Big[ \alpha R_{\mu \nu} R^{\mu \nu} - \beta R^2 +  \kappa^{-2} R \Big].
\ee
This theory is 
renormalizable but unfortunately contains a physical 
ghost
(state of negative norm), implying 
a violation of unitarity in the theory: probability, described by the scattering $S$-matrix, is no more conserved.
The classical theory is unstable, since the dynamics can drive the system to become arbitrarily excited,
and the Hamiltonian constraint is unbounded from below. 

In this paper we generalize the Stelle theory to restore unitarity. 
This work is inspired by papers about a nonlocal extension of 
gauge theories introduced by Moffat, Cornish and their collaborators in the nineties \cite{Moffat}.
The authors further extended the idea to gravity, having in mind the following logic \cite{MoffatG}. 
%
They considered a modification of the Feynman rules 
where the coupling constants ($g_i$ for electro-weak interactions and $G_N$ for gravity)
are no longer constant but function of the momentum $p$. They checked the 
gauge invariance at all orders in gauge theory but only 
up to the second 
order in gravity.
For particular choices of $g_i(p)$ or $G_{N}(p)$, the propagators do not show any other pole  
above the standard particle content of the theory, therefore the theory is unitary.
On the other hand the theory is also finite if the coupling constants go sufficiently fast to zero 
in the ultra-violet limit. 
The problem with gravity is to find a covariant action that self-contains the properties mentioned 
before: finiteness and/or renormalizability and unitarity. 

In the rest of the paper we will mainly refer to the Tomboulis paper \cite{Tombo} that we have 
discovered only recently.
The theory developed in \cite{Tombo}  is very interesting, of great generality and mainly concentrated on 
gauge theories but in \cite{Tombo} the author also concludes with an extension of the idea to gravity.
Here we will use the same notations as standard definitions. 
We arrive to the same conclusions in a different way, but the results are exactly the same. 

The action we are going to consider is a generalization of Stelle's theory
\be 
\hspace{-0.1cm}
 S = 
 - \int \hspace{-0.15cm} 
 \sqrt{ - g} \{ 
 R_{\mu \nu} \alpha (\Box_{\Lambda})  R^{\mu \nu} -  R  \beta (\Box_{\Lambda})  R
+ \gamma \kappa^{-2} R \}  ,
\label{general}
\ee 
where $\alpha(\Box_{\Lambda})$ and 
$\beta(\Box_{\Lambda})$ are now fixed functionals of 
the covariant D'Alembertian operator $\Box_{\Lambda} = \Box/\Lambda^2$ 
and $\Lambda$ is a mass scale in the theory.

In the remainder of this section we will summarize the steps and motivations that led us to the 
generalization (\ref{general}) and will conclude with an extended version 
(\ref{general}).

The covariant action (\ref{general}) 
is a collection of terms, but the 
initial motivation when we started this project was the non local Barvinsky action 
\cite{Barvi}
\begin{eqnarray}
S = - \frac{1}{\kappa^2} \int {\rm d}^4 x \sqrt{ - g} \, 
 G^{\mu \nu}
 \, \frac{1 }{\Box} R_{\mu \nu}, 
\label{theoryBar}
\end{eqnarray}
where $G_{\mu \nu} = R_{\mu \nu} - R \,  g_{\mu \nu}/2$ is the Einstein tensor.  
The action (\ref{theoryBar}) looks like a non-local action, but indeed reproduces the Einstein gravity 
at the lower order in the curvature. Barvinsky has shown that the variation of (\ref{theoryBar}), with 
respect to the metric, gives the following equations of motion (see later in this section for details)
\begin{eqnarray}
R_{\mu \nu} - \frac{1}{2} g_{\mu \nu} R + O(R_{\mu \nu}^2) = 0.
\label{eqBar}
\end{eqnarray} 
The next step is to modify (\ref{theoryBar}) by introducing 
an extra operator 
in the action between the Einstein tensor and the Ricci tensor in order 
to have a well-defined theory at the quantum level 
without loss of covariance.

We consider as guideline the value of the spectral dimension in the definition of a 
finite and/or renormalizable theory of quantum gravity 
which, of course, 
is compatible with the Einstein gravity at low energy 
but also manifests a natural dimensional reduction at high energy.
The dimensional reduction is of primary importance in order to have 
a quantum theory free of divergences in the ultra violet regime.  
The Stelle theory 
is characterized by a two dimensional behavior at high energy 
but this is not sufficient to have a well defined theory at the quantum level because of 
the presence of ghosts in the propagator. 
A first easy generalization of (\ref{theoryBar}) is the following action 
\begin{eqnarray}
S = - \frac{1}{\kappa^2} \int {\rm d}^4 x \sqrt{ - g} \, 
 G^{\mu \nu} F(\Box/\Lambda^2) R_{\mu \nu},
\label{theoryG}
\end{eqnarray}
where $F(\Box/\Lambda^2)$ is a generic function of the 
covariant d'Alambertian operator which satisfies the already mentioned properties
that we are going to summarize below.
\begin{enumerate}
\renewcommand{\theenumi}{(\Roman{enumi})}
\renewcommand{\labelenumi}{\theenumi}
\item Classical limit :
\be
\lim_{\Lambda \rightarrow + \infty} F(\Box/\Lambda^2) = \frac{1}{\Box},
\label{F}
\ee
if the limit is satisfied, the equations of motion are the Einstein equations plus corrections 
in $R_{\mu \nu}^2$ (we will explain this in more detail later in the paper).
\item Finiteness and/or renormalizability of the quantized theory. 
Our guiding principle is to find a well defined quantum theory 
and the dimensional reduction of the spacetime at high energy.
The Stelle theory, the Crane-Smolin theory \cite{CraneSmolin},  
``asimptotically safe quantum gravity" \cite{Reuter}, ``causal dynamical triangulation" \cite{CDT},``loop quantum gravity" \cite{ModeNico} and "string theory" already 
manifest this property with a high energy spectral dimension $d_s = 2$ \cite{Carlip}. 
However 
such reduction is insufficient if we want a unitary theory free from negative norm states.
We can anticipate that,  
for the model that we are going to introduce in this paper, the spectral dimension is smaller than one 
in the ultraviolet regime. 
\end{enumerate}

The general theory (\ref{theoryG}) 
was for the first time derived by Barvinsky \cite{Barvi} in 
the brane-world scenery and can be 
written in the following equivalent way 
\begin{eqnarray}
S = - \frac{1}{\kappa^2} \int {\rm d}^4 x \sqrt{ - g} \, 
 G^{\mu \nu}
 \, \frac{\mathcal{O}^{-1} ( \Box _{ \Lambda} )}{\Box} R_{\mu \nu}.
\label{theory}
\end{eqnarray}
In \cite{Barvi} the author was interested in the infra-red modifications to Einstein gravity. 
In the present paper, however,  
we are interested to ultra-violet modifications to gravity, as stressed in (\ref{F}). 
The form (\ref{theory}) 
of the action is particularly useful to highlight the classical limit and in the 
expression of the equations of motion.
In analogy to the properties 
satisfied by the operator $F(\Box/\Lambda^2)$, 
%
the operator $\mathcal{O}$ has to satisfy the following limit in order to reproduce the classical theory
(see the discussion related to formulas (\ref{theoryBar}) and (\ref{eqBar}))
\begin{eqnarray}
\lim_{\Lambda \rightarrow + \infty} \mathcal{O} (\Box/\Lambda^2) = 1.
\label{cllim}
\end{eqnarray}
 We now add some details about the Barvinsky derivation of the equation of motion.
 Taking the variation of (\ref{theory}) with respect to the metric we find \cite{Barvi}
 \be
 \hspace{-0.5cm}
 - \frac{2}{\kappa^2} \int {\rm d}^4 x \sqrt{-g} \, G^{\mu \nu} \frac{\mathcal{O}^{-1} ( \Box _{ \Lambda} )}{\Box} \, \delta_g R_{\mu \nu} + O(R_{\mu \nu}^2). 
 \ee
 Since 
 $\delta_g R_{\mu \nu} = -\frac{1}{2} \Box \delta g_{\mu \nu} + \nabla_{\mu} \epsilon_{\nu} + \nabla_{\nu} \epsilon_{\mu}$ \cite{Barvi}, 
 integration by parts cancels the $\Box$ operator at the denominator 
 and the contribution of the gauge parameters $\epsilon_{\mu}$ vanishes in view of the
 Bianchi identities, $\nabla^{\mu} G_{\mu \nu} = 0$.  
 All the commutators of covariant derivatives with the $\Box$ operator in
 $\mathcal{O}^{-1} ( \Box _{ \Lambda} )/\Box$ give rise to curvature square operators. 
 Also, the direct variation of the metric gives rise to curvature square terms.
 Then the equations of motion are very simple, 
if we omit the 
squared curvature terms: 
\begin{eqnarray}
\hspace{-0.3cm} 
\mathcal{O}^{-1} \left( \Box_{\Lambda} \right) \left( R_{\mu \nu} -\frac{1}{2} g_{\mu \nu} R \right)  
+ O(R_{\mu \nu}^2) 
= 
8 \pi G_N T_{\mu \nu}. 
\label{newgravity0}
\end{eqnarray}
If we truncate the theory to the linear part in the Ricci curvature, the equations of motions simplify to  
\begin{eqnarray}
&& R_{\mu \nu} -\frac{1}{2} g_{\mu \nu} R = 
8 \pi G_N \mathcal{O}( \Box_{ \Lambda} ) T_{\mu \nu}, \nonumber \\
&& \nabla^{\mu} \left(  \mathcal{O}( \Box_{\Lambda} ) T_{\mu \nu} \right) = 0,
\label{newgravity}
\end{eqnarray}
where the second relation is a consequence of the Bianchi identities that we impose on the solution.
In particular, such relation implies that the conserved quantity has to be
${\cal S}_{\mu \nu} = \mathcal{O}( \Box / \Lambda^2 ) T_{\mu \nu}$
in the 
truncation of the theory. These equations of motion are very interesting because black hole solutions,
at least for a particular choice of the operator $\mathcal{O}$, are not singular anymore, as recently showed in \cite{ModestoMoffatNico}.

At the classical level, equations (\ref{newgravity0}) can be 
derived from the action (\ref{theoryG}) or (\ref{theory}) but 
at the quantum level also the Einstein-Hilbert action and the high derivative terms 
introduced by Stelle are generated. Thus, the complete Lagrangian we are going to study is
\be
&& \hspace{-0.4cm}
\mathcal{L} =   - \sqrt{ - g} \, \Big[ \frac{\beta}{\kappa^2} R - \beta_2 \left(R_{\mu \nu} R^{\mu \nu} 
- \frac{1}{3} R^2 \right)
+ \beta_0 R^2 \nonumber \\
&& \hspace{-0.4cm}
+ \left(R_{\mu \nu} \, h_2 \left( - \Box_{\Lambda} \right) \, R^{\mu \nu} 
- \frac{1}{3} R \,h_2 \left( - \Box_{\Lambda} \right) \, R \right) \label{final}
\\
&&  \hspace{-0.4cm}
- R \, h_0 \left( - \Box_{\Lambda} \right)  R  
-\frac{1}{2 \xi} F^{\mu}  \omega( - \Box_{\Lambda}^{\eta})  F_{\mu} + 
\bar{C}^{\mu} M_{\mu \nu} C^{\nu} \Big].
\nonumber 
\ee
The operator $\Box^{\eta}_{\Lambda}$ encapsulates the D'Alembertian of the flat fixed background,
whereas  
$F_{\mu}$ is the gauge fixing function with the weight functional $\omega$.
In general we introduce two different functions $h_2$ and $h_0$. Those functions have not to be 
polynomial but \emph{entire functions without poles or essential singularities}. 
While nonlocal kernels can lead to unitary problems, 
the functions $h_2$ and $h_0$ do introduce an effective non-locality. 
However, since $h_2$ and $h_0$ are transcendental entire
functions, their behavior is quite similar to polynomial functions and 
unitary problems do not occur. 


Let us assume for a moment that $h_i(x) = p_n(x)$, where $p_n(x)$ is a polynomial of degree $n$. 
In this case, as it will be evident in the next section, the propagator takes the following 
form 
\be
\frac{1}{k^2(1+ p_n(k^2))} = \frac{c_0}{k^2} + \sum_i \frac{c_i}{k^2 - M_i^2},
\label{poli}
\ee
\vspace{0.1cm}\\
where we used the factorization theorem for polynomial and the partial fraction 
decomposition \cite{Tombo}. When multiplying by $k^2$ the left and right side of (\ref{poli}) and 
considering 
the ultraviolet limit, we find that at least one of the coefficients $c_i$ is negative, therefore 
the theory contains a ghost 
in the spectrum. 
The conclusion is that $h_2$ and $h_0$ can not be polynomial.

We can also add to the action the Kretschmann scalar 
$R_{\mu\nu\rho\sigma} R^{\mu\nu\rho\sigma}$ but, for spacetimes topologically equivalent 
to the flat space, we can use the Gauss-Bonnet topological invariant 
\be
\int d^4 x \sqrt{-g} [ R_{\mu\nu\rho\sigma} R^{\mu\nu\rho\sigma} - 4 R_{\mu \nu} R^{\mu \nu} + R^2]
\ee 
to rephrase the Kretschmann invariant in terms of $R^2$ and $R_{\mu\nu} R^{\mu \nu}$ 
already present in the action. 
%
%
%
%
\section{BRST invariance and Graviton propagator}
We start by considering the quadratic expansion of the Lagrangian (\ref{final}) in the graviton field
fluctuation without specifying the explicit form of the functionals $h_2$ and $h_0$ (if not necessary).
Following the Stelle paper, we expand 
around the Minkowski background $\eta^{\mu \nu}$ 
in power of the graviton field $h^{\mu \nu}$  defined in the following way  
\begin{eqnarray}
\sqrt{ - g} g^{\mu \nu} = \eta^{\mu \nu} + \kappa h^{\mu \nu}. 
\label{graviton}
\end{eqnarray}
The form of the propagator depends not only on the gauge choice but also on the definition of the 
gravitational fluctuation \cite{Shapirobook}. 
In the quantum theory 
the gauge choice is the familiar {\em harmonic gauge} 
$\partial_{\nu} h^{\mu \nu} =0$
and the Green's functions are defined by the generating functional 
\begin{widetext}
\begin{centering}  
\be
Z(T_{\mu\nu}) = \mathcal{N} \int \prod_{\mu\leqslant \nu} d h^{\mu\nu} [d C^{\sigma}] 
[d \bar{C}_{\rho}] [d e^{\tau}] \, \delta(F^{\tau} - e^{\tau}) 
e^{ i \big\{ S_g - \frac{1}{2 \xi} \int d^4 x  \,e_{\tau} \, \omega(- \Box^{\eta}_{\Lambda}) \, e^{\tau} 
+ \int d^4 x \, \bar{C}_{\tau} F^{\tau}_{\mu \nu} D^{\mu \nu}_{\alpha} C^{\alpha}
+ \kappa \int d^4 x \, T_{\mu \nu} h^{\mu \nu}
\big\} } , \label{Z} 
\ee
\end{centering}
\end{widetext}
where $S_g$ is the gravitational action defined in (\ref{final}) subtracted of the gauge and ghost terms and
$F^{\tau} = F^{\tau}_{\mu \nu} h^{\mu \nu}$ with $F^{\tau}_{\mu \nu} = \delta^{\tau}_{\mu}\partial_{\nu}$. 
$D^{\mu \nu}_{\alpha}$ is the operator which generates the gauge transformations in the graviton 
fluctuation $h^{\mu \nu}$. Given the infinitesimal coordinates transformation 
$x^{\mu \prime} = x^{\mu} + \kappa \xi^{\mu},$ 
the graviton field transforms as follows 
\be
&& \delta h^{\mu \nu} = D^{\mu \nu}_{\alpha} \xi^{\alpha} 
= \partial^{\mu} \xi^{\nu} + \partial^{\nu} \xi^{\mu} - \eta^{\mu \nu}  \partial_{\alpha} \xi^{\alpha} \nonumber \\
&& \hspace{-0.5cm}
+ \kappa ( \partial_{\alpha} \xi^{\mu} h^{\alpha \nu} + \partial_{\alpha} \xi^{\nu} h^{\alpha \mu}
- \xi^{\alpha} \partial_{\alpha} h^{\mu \nu} - \partial_{\alpha} \xi^{\alpha} h^{\mu \nu} ).
\ee
We have also introduced a weighting gauge functional (to be precise the second term in (\ref{Z})),  
which depends on a weight function 
$\omega( - \Box_{\Lambda})$ with the property to fall off at least like the entire functions $h_2(k^2/\Lambda^2)$, $h_0(k^2/\Lambda^2)$
for large momenta \cite{Stelle}. 

When the gauge symmetry is broken by the addition of the gauge-fixing term, 
a residual transformation survives for the effective action which involves the gravitational,
gauge-fixing and ghost actions terms. This is the BRST symmetry defined by the following 
transformation, which is appropriate for the gauge-fixing term,
\be
&& \delta_{\rm BRST} h^{\mu \nu} =  \kappa D^{\mu \nu}_{\alpha} C^{\alpha} \delta \lambda,  \nonumber \\
&& \delta_{\rm BRST} C^{\alpha} = - \kappa^2 \partial_{\beta} C^{\alpha} C^{\beta} \delta \lambda, \nonumber \\
&& \delta_{\rm BRST} \bar{C}^{\alpha} = - \kappa \xi^{-1} \omega(-\Box^{\eta}_{\Lambda}) F^{\tau} \delta \lambda,
\label{BRST}
\ee
where $\delta \lambda$ is a constant infinitesimal anticommuting parameter.
The first transformation in (\ref{BRST}) is nothing but a gauge transformation generated by 
$\kappa C^{\alpha} \delta \lambda$, so the functional $S_g$ is BRST invariant since it is 
a function of $h^{\mu \nu}$ alone. Other two BRST invariant quantities are 
\be
&& \delta_{\rm BRST}^2 C^{\alpha} \sim \delta_{\rm BRST} ( \partial_{\beta} C^{\sigma} C^{\beta} ) =0, \nonumber \\
&& \delta_{\rm BRST}^2 h^{\mu \nu}  \sim \delta_{\rm BRST} ( D^{\mu \nu}_{\alpha} C^{\alpha} ) =0.
\label{nilh}
\ee
The above transformation follows from the anticommuting nature of $C^{\alpha}$, $\delta \lambda$
and the following commutation relation of two gauge transformations generated by $\xi^{\mu}$ and
$\eta^{\mu}$,
\be
\hspace{-0.2cm}
\frac{\delta D^{\mu \nu}_{\alpha}}{\delta h^{\rho \sigma}} D^{\rho \sigma} 
(\xi^{\alpha} \eta^{\beta} - \eta^{\alpha} \xi^{\beta} ) 
\hspace{-0.08cm} 
=
\hspace{-0.08cm}
 \kappa D^{\mu \nu}_{\lambda} 
(\partial_{\alpha} \xi^{\lambda} \eta^{\alpha} - \partial_{\alpha} \, \eta^{\lambda} \, \xi^{\alpha}).
\ee
Given the second of (\ref{nilh}), only the antighost $\bar{C}_{\tau}$ transforms under the BRST transformation to cancel the variation of the gauge-fixing term. The entire effective action is BRST invariant 
\be
&&  \hspace{-0.0cm}
 \delta_{\rm BRST} 
\Big( {S}_g - \frac{1}{2 \xi} \int 
d^4 x \, 
F_{\tau} 
\omega(-\Box^{\eta}_{\Lambda}) F^{\tau} \nonumber \\
&& \hspace{1.2cm} 
+  \int d^4 x \, 
\bar{C}_{\tau} F^{\tau}_{\mu \nu} D^{\mu \nu}_{\alpha} C^{\alpha} \Big) =0. 
\ee
Let us list the mass dimension of the fields in the gauge-fixed Lagrangian:
$[h^{\mu \nu}] ={\rm mass}$, $[C^{\tau}] ={\rm mass}$, 
$[\bar{C}_{\tau}] ={\rm mass}$, $[\kappa]={\rm mass}^{-1}$.

Now we Taylor-expand the gravitational part of the action (\ref{theory}) to the second order 
in the gravitational perturbation $h^{\mu \nu}(x)$ to obtain the graviton propagator. 
In the momentum space, the action which is purely quadratic in the gravitational field, 
reads  
\begin{widetext}
\begin{centering}  
\begin{eqnarray}
&& \hspace{0cm} 
S^{(2)} = \int 
\hspace{-0.1cm}
d^4 k \, h^{\mu \nu} (-k) 
 \Big( - [ \beta - \beta_2 \kappa^2 k^2 + \kappa^2 k^2 h_2(k^2/\Lambda^2) ] \, k^2 \, P^{(2)}_{\mu \nu \rho \sigma}(k)
+ \xi^{-1} 
\,  \omega(k^2/\Lambda^2)  P^{(1)}_{\mu \nu \rho \sigma}(k)
  \nonumber \\
 && 
\hspace{1.0cm} 
+ \{ 3 \, k^2 \,  [ \beta/2 -3 \beta_0 \kappa^2 k^2 + 3 \kappa^2 k^2 h_0(k^2/\Lambda^2) ] 
+ 2 \xi^{-1} \omega(k^2/\Lambda^2) \} P^{(0 - \omega)}_{\mu \nu \rho \sigma}(k) \nonumber \\
&& \hspace{1.0cm}
+ k^2 \,  [ \beta/2 - 3 \beta_0 \kappa^2 k^2 + 3 \kappa^2 k^2  h_0(k^2/\Lambda^2) ] 
\{   P^{(0 - s)}_{\mu \nu \rho \sigma}(k) + \sqrt{3} [ P^{(0 - \omega s)}_{\mu \nu \rho \sigma}(k)
+ P^{(0 - s \omega)}_{\mu \nu \rho \sigma}(k) ] \} \Big) h^{\rho \sigma}(k), 
\label{quadratic}
 %
 %
\end{eqnarray}
\end{centering}
\end{widetext}
where we used the gauge $F^{\tau}= \partial_{\mu} h^{\mu \tau}$ and where we have introduced the 
projectors \cite{VN}
\be
 && P^{(2)}_{\mu \nu \rho \sigma}(k) = \frac{1}{2} ( \theta_{\mu \rho} \theta_{\nu \sigma} +
 \theta_{\mu \sigma} \theta_{\nu \rho} ) - \frac{1}{3} \theta_{\mu \nu} \theta_{\rho \sigma}  ,
 \nonumber \\
 && P^{(1)}_{\mu \nu \rho \sigma}(k) = \frac{1}{2} \left( \theta_{\mu \rho} \omega_{\nu \sigma} +
 \theta_{\mu \sigma} \omega_{\nu \rho}  + 
 \theta_{\nu \rho} \omega_{\mu \sigma}  +
  \theta_{\nu \sigma} \omega_{\mu \rho}  \right) ,
 \nonumber \\
 && P^{(0 - s)} _{\mu\nu\rho\sigma} (k) = \frac{1}{3}  \theta_{\mu \nu} \theta_{\rho \sigma} \,\, , \hspace{0.1cm}
P^{(0 - \omega)} _{\mu\nu\rho\sigma} (k) =  \omega_{\mu \nu} \omega_{\rho \sigma}, \nonumber \\
&& P^{(0 - s \omega)} _{\mu\nu\rho\sigma}  = \frac{1}{\sqrt{3}}  \theta_{\mu \nu} \omega_{\rho \sigma}, 
\,\,
 \hspace{0.1cm} P^{(0 - \omega s )} _{\mu\nu\rho\sigma} =  \frac{1}{\sqrt{3}} \omega_{\mu \nu} \theta_{\rho \sigma}, \label{proje}
\ee
\vspace{1cm}
where we defined the transverse and longitudinal projectors for vector quantities 
\be
 \theta_{\mu \nu} = \eta_{\mu \nu} - \frac{k_{\mu} k_{\nu}}{k^2} \,\, , \hspace{0.1cm}
 \omega_{\mu \nu} = \frac{k_{\mu} k_{\nu}}{k^2}. 
\ee
\vspace{1cm}
Using the orthogonality properties of (\ref{proje}) we can now invert the kinetic matrix in 
(\ref{quadratic}) and obtain the 
graviton propagator. In the following expression the graviton propagator is expressed 
in the momentum space according to the quadratic action (\ref{quadratic}), 
\begin{widetext}
\begin{centering}  
\be
&& D_{\mu\nu\rho\sigma}(k) = \frac{- i}{(2 \pi)^4} \frac{1}{k^2 + i \epsilon} 
\Bigg( \frac{2 P^{(2)}_{\mu \nu \rho \sigma}(k)}{\beta - \beta_2 \kappa^2 k^2 + \kappa^2 k^2 h_2(k^2/\Lambda^2)}
- \frac{4 P^{(0 - s)}_{\mu \nu \rho \sigma}(k) }{ \beta - 6 \beta_0 \kappa^2 k^2 + \kappa^2 k^2 h_0(k^2/\Lambda^2)} \nonumber \\
&& \hspace{2cm}
- \frac{2 \xi P^{(1)}_{\mu \nu \rho \sigma}(k)}{\omega(k^2/\Lambda^2)}
 -\xi \frac{ 3 P^{(0 - s)}_{\mu \nu \rho \sigma}(k) - \sqrt{3} [ P^{(0 - s \omega)} _{\mu\nu\rho\sigma} (k) + 
P^{(0 - \omega s )} _{\mu\nu\rho\sigma} (k) ] + P^{(0 -  \omega)} _{\mu\nu\rho\sigma} (k)
}{\omega(k^2/\Lambda^2)} \Bigg). \label{compPro}
\ee
\end{centering} 
\end{widetext}
Let us consider the graviton propagator in the gauge $\xi = 0$. In this  particular gauge, only 
the first two terms in (\ref{compPro}) survive. 
We will show in the next section that only the physical massless spin-2 pole occurs in the 
propagator
when the theory is renormalized at a certain scale $\mu_0$. The renormalization group invariance 
preserves unitarity in the dressed physical propagator at any energy scale and no other 
physical pole emerges at any other scale.

\section{ 
 Renormalizability}

In this section we want to find an upper bound to the divergences in quantum gravity; before doing 
this, we have to construct the entire functions $h_2$ and $h_0$.
Looking at the first two gauge invariant terms in (\ref{compPro}), we introduce the following notation 
\be
&& \bar{h}_2(z) = \beta - \beta_2 \kappa^2 \Lambda^2 z + \kappa^2 \Lambda^2 z \, h_2(z) , \nonumber \\
&& \bar{h}_0(z) = \beta - 6 \beta_0 \kappa^2 \Lambda^2 z + 6 \kappa^2 \Lambda^2 z \, h_0(z), 
\ee
where $z$ will be identified with $ - \Box_{\Lambda}$. 

Considering \cite{Tombo}, we require the following general properties for the transcendental entire functions $h_i$ ($i = 2,0$):
\begin{enumerate}
\renewcommand{\theenumi}{(\roman{enumi})}
\renewcommand{\labelenumi}{\theenumi}
\item $\bar{h}_i(z)$ is real and positive on the real axis, it has no zeroes on the 
whole complex plane $|z| < + \infty$. This requirement implies that there are no 
gauge-invariant poles other than the transverse massless physical graviton pole.
\item $|h_i(z)|$ has the same asymptotic behavior along the real axis at $\pm \infty$.
\item There exists $\Theta>0$ such that 
$$\lim_{|z|\rightarrow + \infty} |h_i(z)| \rightarrow | z |^{\gamma} \,\, , \,\,\,\,  
\gamma\geqslant 2$$ 
for the argument of $z$ in the cones 
\be
&& \hspace{-0.2cm} 
C = \{ z \, | \,\, - \Theta < {\rm arg} z < + \Theta \, , \,\,  \pi - \Theta < {\rm arg} z < \pi + \Theta \} , \nonumber \\
&&  \hspace{-0.2cm} 
{\rm for } \,\,\, 0< \Theta < \pi/2. \nonumber 
\ee
This condition is necessary in order to achieve the supe-renormalizability of the theory. The necessary 
asymptotic behavior is imposed not only on the real axis, (ii) but also in conic regions surrounding 
the real axis. In an Euclidean spacetime, the condition (ii) is not strictly necessary if (iii) applies.
\end{enumerate}
Given the above properties, let us study the ultraviolet behavior of the quantum theory.
From the property (iii) in the high energy regime, the propagator in momentum space goes as 
$1/k^{2 \gamma +4}$ (see (\ref{compPro})), but also the $n$-graviton interaction has the 
same scaling, since it can be written in the following schematic way,
\be
&& \hspace{-0.25cm}
{\mathcal L}^{(n)} \sim  h^n \, \Box_{\eta} h \,\,  h_i( - \Box_{\Lambda}) \,\, \Box_{\eta} h \nonumber \\
&& \hspace{0.5cm}
\rightarrow h^n \, \Box_{\eta} h 
\,  ( \Box_{\eta} + h^m \, \partial h \partial )^{\gamma} \, 
\Box_{\eta} h , 
\label{intera}
\ee
where $h$ is the graviton field and $h_i$ is the entire function defined by the properties (i)-(iii). 
From (\ref{intera}), the superficial degree of divergence (in four spacetime dimensions) is 
\be
\hspace{-0.3cm} 
D = 4 L - (2 \gamma + 4) I + ( 2 \gamma + 4) V 
= 4 - 2 \gamma (L - 1).
\label{diver}
\ee
In (\ref{diver}) we used again the topological relation between vertexes $V$, internal lines $I$ and 
number of loops $L$: $I = V + L -1$. 
Thus, if $\gamma \geqslant 3$, only 1-loop divergences exist and the theory is super-renormalizable.
In this theory 
 the quantities 
$\beta$, $\beta_2$, $\beta_0$ and eventually the cosmological constant are renormalized, 
\be
&& \hspace{-0.7cm}
\mathcal{L}_{\rm Ren} = \mathcal{L}  - 
\sqrt{-g} 
\Big\{  \frac{\beta(Z -1)}{\kappa^2} R  + \lambda (Z_{\lambda} - 1) \nonumber \\
&& \hspace{-0.7cm}
- \beta_2 (Z_2 - 1) (R_{\mu \nu} R^{\mu \nu} - \frac{1}{3} R^2 )
+ \beta_0 ( Z_0 -1 ) R^2 \Big\} ,
\ee
where all the coupling must be understood as renormalized at an energy scale $\mu$. On the other hand,  
the functions $h_i$ are not renormalized. In order to better understand this point we can write 
the generic entire functions as series, $h_i(z) = \sum_{r=0}^{+\infty} a_r z^r$.
For $r \geqslant 1$ there are no counterterms that renormalize $a_r$ 
because of the superficial degree of divergence (\ref{diver}). Only the 
coefficient $a_0$ is renormalized but this is just a 
normalization convention. 
The non-trivial dependence of the entire functions $h_i$ on their argument is preserved at quantum level.

Imposing the conditions (i)-(iii) we have the freedom to choose the following form for the functions 
$h_i$,
\be
&& h_2(z) = \frac{\bar{h}_2 (z) - \alpha + \alpha_2 z}{\kappa^2 \Lambda^2 z}, \nonumber \\
&& h_0(z) = \frac{\bar{h}_0 (z) - \alpha + \alpha_0 z}{6 \kappa^2 \Lambda^2 z},
\label{hz}
\ee
for three general parameters $\alpha$, $\alpha_2$ and $\alpha_0$. 
In general in order to be compatible with the conditions (i)-(iii), we have to take $\bar{h}_i(z) = \alpha \, \exp H_i(z)$, where
$H_i(z)$ is an entire function that exhibits logarithmic asymptotic behavior in the conical region $C$.
Since $H(z)$ is an entire function, $\bar{h}_i(z)$ has no zeros in all complex plane for $|z|< + \infty$.
Furthermore, the non-locality in the action is actually a kind of non locality, because $\bar{h}_i(z)$ is an exponential 
function and a Taylor expansion of $h_i(z)$ erases the denominator $\Box_{\Lambda}$ at any energy scale.
For this reason, the entire functions $h_i(z)$ can be called {\em quasi polynomial}.

The entire function $H(z)$ which satisfies the property (iii), 
can be defined as 
\be
H(z) = \int_0^{p_{\gamma +1 }(z)} \frac{1 - \zeta(\omega)}{\omega} {\rm d} \omega \, , 
\label{Hz}
\ee
where the following requirements have to be satisfied :
\begin{enumerate}
\renewcommand{\theenumi}{\alph{enumi}.}
\renewcommand{\labelenumi}{\theenumi}
\item $p_{\gamma +1 }(z)$ is a real polynomial of degree $\gamma+1$ with $p_{\gamma +1}(0) = 0$,
\item $\zeta(z)$ is an entire and real function on the real axis with $\zeta(0) = 1$,
\item $|\zeta(z)| \rightarrow 0$ for $|z| \rightarrow \infty$ in the conical region $C$ defined in (iii). 
\end{enumerate}

Let us assume now that the theory is renormalized at some scale $\mu_0$. 
If we want that the bare propagator to possess no other gauge-invariant pole than 
the transverse physical graviton pole, we have to set  
\be
&&  \alpha = \beta(\mu_0), \nonumber \\
&& \frac{\alpha_2}{\kappa^2 \Lambda^2} =  \beta_2(\mu_0), \nonumber \\
&&   \frac{\alpha_0}{6 \kappa^2 \Lambda^2} = \beta_0(\mu_0) .
\label{betaalpha}
\ee
As pointed out in \cite{Tombo}, the relations (\ref{betaalpha}) can be used to fix the introduced scale 
$\Lambda$ in terms of the Planck scale $\kappa^{-2}$.
If we fix 
\be
&& \beta(\mu_0) = \alpha , \nonumber \\
&& \frac{\beta_2 (\mu_0)}{\alpha_2} = \frac{ 6 \beta_0 (\mu_0)}{\alpha_0}, 
\ee
the two mass scales are linked by the following relation 
\be
\Lambda^2 = \frac{\alpha_2}{\kappa^2 \beta_2(\mu_0)}.
\ee 

If the energy scale $\mu_0$ is taken as the renormalization point we find 
$\bar{h}_2 = \bar{h}_0 = \beta(\mu_0) \, \exp H(z):=\bar{h}(z)$ and only the physical massless spin-2 graviton pole
occurs in the bare propagator. In the gauge $\xi = 0$ the propagator in (\ref{compPro}) 
simplifies to 
\be 
&& \hspace{-0.5cm} 
D_{\mu\nu\rho\sigma}(k) = \frac{- i}{(2 \pi)^4} \frac{1}{k^2 + i \epsilon} 
\Bigg( \frac{ 2 P^{(2)}_{\mu \nu \rho \sigma}(k)
- 4 P^{(0)}_{\mu \nu \rho \sigma}(k)}{\alpha \, \bar{h}(k^2/\Lambda^2) }\Bigg)  \nonumber \\
&& \hspace{0.0cm} 
= \frac{- i}{(2 \pi)^4} \frac{e^{- H(k^2/\Lambda^2)}}{\alpha \, ( k^2 + i \epsilon) } 
\Big(  2 P^{(2)}_{\mu \nu \rho \sigma}(k)
- 4 P^{(0)}_{\mu \nu \rho \sigma}(k) \Big) .
\label{propalpha}
\ee
If we choose another renormalization scale $\mu$, then the bare propagator acquires poles; 
however, these poles 
cancel in the dressed physical propagator because the shift in the bare part is cancelled 
with a corresponding shift in the self energy. This follows easily from the renormalization group invariance.
The same procedure is not applicable to the case $h_2(z)=h_0(z)=0$ \cite{Stelle}, because 
the theory fails to be renormalizable when the unitarity requirement $\beta_2=\beta_0=0$ is imposed
and an infinite tower of counterterms has to be added to the action.

An explicit example of $\bar{h}(z)=\exp H(z)$ that satisfies the properties (i)-(iii) can be easily constructed. 
There are of course many ways to choose $\zeta(z)$, but we focus here on the obvious exponential choice  
$\zeta(z) = \exp(- z^2)$, which satisfies property c. in a conical region $C$ with 
$\Theta =\pi/4$. 
The entire function $H(z)$ is the result of the integral defined in (\ref{Hz})
\be
&& 
H(z) = \frac{1}{2} \left[ \gamma_E + 
\Gamma \left(0, p_{\gamma+1}^{2}(z) \right) \right] + \log [ p_{\gamma+1}(z) ] \, , 
\nonumber \\
&& {\rm Re}( p_{\gamma+1}^{2}(z) ) > 0, 
\ee
where $\gamma_E=0.577216$ is the Euler's constant and  
$\Gamma(a,z) = \int_z^{+ \infty} t^{a -1} e^{-t} \rm{d} t$ is the incomplete gamma function.  
If we choose $p_{\gamma+1}(z) = z^{\gamma +1}$, $H(z)$ simplifies to:
\be
&& 
H(z) = \frac{1}{2} \left[ \gamma_E + \Gamma \left(0, z^{2 \gamma +2} \right) \right] + \log (z^{\gamma +1}) \, ,
\nonumber \\
&& {\rm Re}(z^{2 \gamma +2}) > 0. 
\label{H0}
\ee
Another equivalent expression for the entire function $H(z)$ is given by the following series
\be
&& H(z) = \sum_{n =1}^{+ \infty} ( -1 )^{n-1} \, \frac{p_{\gamma +1}(z)^{2 n}}{2n \, n!},
\nonumber \\
&& {\rm Re}( p_{\gamma+1}^{2}(z) ) > 0.
\label{HS}
\ee
\begin{figure}[ht]
\begin{center}
\hspace{-0.7cm}
\includegraphics[width=4.32cm,angle=0]{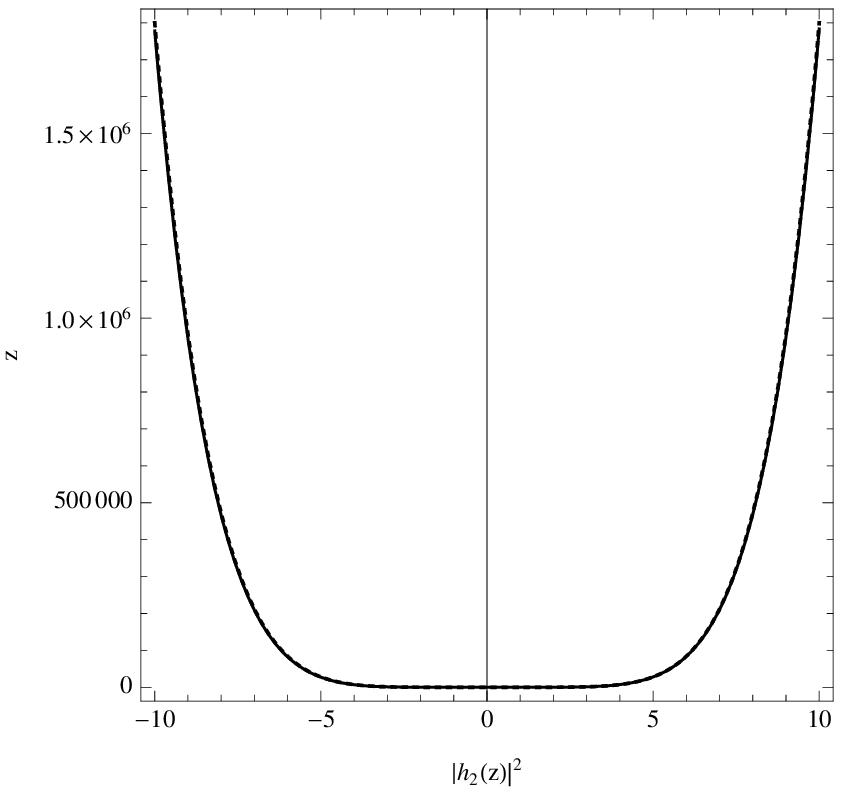}
\includegraphics[width=4cm,angle=0]{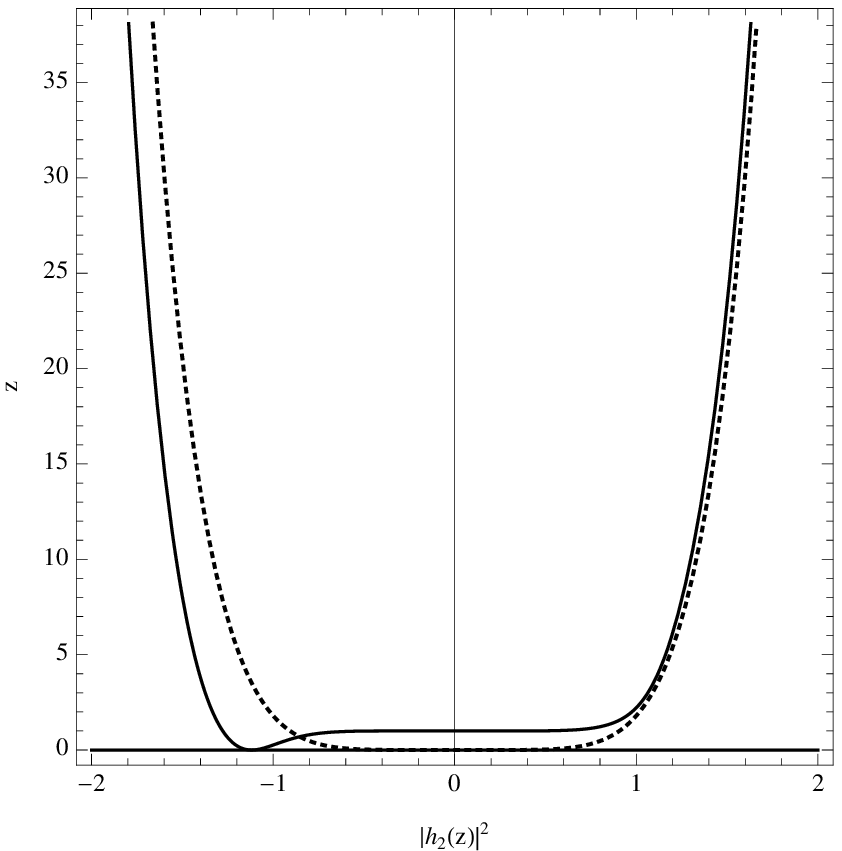}
\caption{\label{plotrhoeff} Plot of $|h(z)|^2$ for $z$ real and $\alpha = \alpha_2 =1$ (solid line). 
In the first plot $z\in[-10,10]$ and in the second one $z\in [-2,2]$. 
The dashed line represents the asymptotic limit for large real positive and negative values of $z$.
The asymptotic behavior is $ |h(z)|^2 \approx 1.8 \, z^6$.}
\end{center}
\end{figure}
For $p_{\gamma +1}(z) = z^{\gamma +1}$ the $\Theta$ angle, which defines the cone $C$,  
is $\Theta = \pi/(4 \gamma+4)$. 
According to the above expression (\ref{HS}) we find the following behavior near $z = 0$ for the 
particular choice $p_{\gamma+1}(z) = z^{\gamma +1}$,
\be
H(z) = \frac{ z^{2 \gamma + 2}}{2} - \frac{ z^{4 \gamma + 4}}{8} + 
\frac{ z^{6 \gamma + 6}}{36} + O(z^{6 \gamma + 7}),
\ee
where the Taylor expansion is the exact one 
for ${\rm Arg}(z) < \pi/8$
but we already have a stronger constraint on the cone $C$, 
${\rm Arg}(z) < \Theta = \pi/(4 \gamma+4)$. In particular 
$ \lim_{z \rightarrow 0} H(z) = 0$\footnote{
We can do a simple choice of the entire function $H(z)$, which gives 
rise to a condition stronger than (iii). If we take $H(z) = z^2$, then 
\be
 \bar{h}_2(z) = \bar{h}_0(z) = \alpha \,  e^{z^2}, 
\label{exps}
\ee
where again $z = - \Box/\Lambda^2$. 
Another possible choice we wish to analyze is $\bar{h}_2 = \bar{h}_0 = \alpha \, e^{z}$,
%
because of its connection with the regular Nicolini-Spallucci black holes \cite{NS}.\\
For the functions given in (\ref{exps}), the upper bound in (\ref{diver}) can be derived as a truncation 
of the exponentials and the result does not change: we have divergences at one loop, but 
the theory is finite for $L > 1$. 
The exponentials in (\ref{exps}) improve the convergence
properties of the theory and the propagator is:
\be 
&& \hspace{-0.65cm} 
D_{\mu\nu\rho\sigma}(k)  
%
= \frac{- i}{(2 \pi)^4} \frac{2 \, e^{- k^4/\Lambda^4}}{\alpha ( k^2 + i \epsilon) } 
\Big(  \hspace{-0.05cm} P^{(2)}_{\mu \nu \rho \sigma}(k)
- 2 P^{(0)}_{\mu \nu \rho \sigma}(k)  \hspace{-0.05cm} \Big) . 
\ee 
For $\bar{h}_i(z) = \alpha \exp(z)$ the exponential in the above propagator is replaced with 
$\exp( - k^2/\Lambda^2)$.}.



\section{Spectral dimension}
In this section, we calculate the spectral dimension of the spacetime at short distances, showing 
that the renormalizability, together with the unitarity of the theory, implies a spectral dimension 
smaller than one.
Let us summarize the definition of spectral dimension in quantum gravity.
The definition of spectral dimension is borrowed
from the theory of diffusion processes on fractals \cite{avra} and easily 
adapted to the
quantum gravity context. 
Let us study the Brownian motion of a test particle moving on a $d$-dimensional Riemannian manifold $\mathcal M$ with a fixed smooth metric $g_{\mu\nu}(x)$.
The probability density for the particle to diffuse from $x'$ to $x$ during the fictitious time 
$T$
 is the heat-kernel $K_g(x,x';T)$, which satisfies the heat equation
\begin{eqnarray}
\label{heateq}
\partial_T K_g(x,x';T)=\Delta_g^{{\rm eff}} K_g(x,x';T)
\end{eqnarray}
where $\Delta_g^{{\rm eff}}$ denotes the effective covariant Laplacian. 
It is the usual covariant Laplacian at low energy but it can undergo strong modification in the
ultra-violet regime.
In particular, we will 
be interested in the effective Laplacian at high energy and in relation to the flat background.
The heat-kernel is a matrix element of the operator 
$\exp(T\,\Delta_g)$, acting on the real Hilbert space $L^2(\mathcal M, \sqrt{g} \, \text{d}^d x)$, between position eigenstates
\begin{eqnarray}
K_g(x,x';T) =\langle x^{\prime} | \exp(T\,\Delta_g^{\rm eff}) | x \rangle.
\label{EK}
\end{eqnarray} 
Its trace per unit volume,
\begin{eqnarray}
\label{trace}
&& P_g(T)\equiv V^{-1}\int \text{d}^dx\,\sqrt{g(x)}\,K_g(x,x;T) \nonumber \\
&& \hspace{1cm} \equiv 
V^{-1}\,{\rm Tr}\,
\exp(T\,\Delta_g^{\rm eff})
\end{eqnarray}
has the interpretation of an average return probability. Here $V\equiv\int
d^dx\,\sqrt{g}$ denotes the total volume. It is well known that $P_g(T)$
possesses an asymptotic expansion for $T\rightarrow 0$ of the form
$P_g(T)=(4\pi T)^{-d/2}\sum_{n=0}^\infty A_n\,T^n$. The coefficients $A_n$ have a geometric meaning, i.e. $A_0$ is the volume of the manifold and, if $d=2$, $A_1$ is then proportional to the Euler characteristic. From the knowledge of the function $P_g(T)$ one can recover the dimensionality of the
manifold as the limit for small $T$ of
\begin{eqnarray}
\label{dimform}
d_s \equiv-2\frac{\text{d}\ln P_g(T)}{\text{d}\ln T}.
\label{defi}
\end{eqnarray}
If we consider arbitrary fictitious times $T$, this quantity might depend on the scale we are probing. 
Formula (\ref{defi}) 
is the definition of fractal dimension we will use.

From the bare graviton propagator (\ref{propalpha}) 
we can easily obtain the heat-kernel and then the spectral 
dimension of the quantum spacetime. 
In short, in the momentum space the graviton propagator, omitting the tensorial structure that 
does not affect the 
spectral dimension, 
reads 
\be
D(k) \propto \frac{1}{k^2 \, \bar{h}(k^2/\Lambda^2)}.
\label{propF}
\ee
We also know that the propagator (in the coordinate space) and the heat-kernel are related by \cite{manualheat}
\be
&& \hspace{-1cm} 
 G(x,x^{\prime}) = \int_0^{+\infty} {\rm d}T\, K_g(x, x^{\prime}; T)  \nonumber \\
&& \hspace{0.3cm} \propto \int {\rm d}^4 k \, e^{i k (x- x^{\prime})} \int_0^{+ \infty}  {\rm d} T \, K_g(k; T) ,
\label{eatK}
\ee
where $G(x, x^{\prime}) \propto \int {\rm d}^4 k \, \exp[i k (x - x^{\prime}) ] D(k)$ is the Fourier transform 
of (\ref{propF}).
Given the propagator (\ref{propF}), 
it is easy to invert (\ref{eatK}) with the heat-kernel in the momentum space, 
\be
K_g(k; T) \propto \exp[ - k^2 \, \bar{h}(k^2/\Lambda^2) \, T \, ]\label{Kk},
\ee 
which is the solution of the heat-kernel equation (\ref{heateq}) with the effective operator 
\be
\Delta^{\rm eff}_g = \bar{h}( - \Delta_g/\Lambda^2) \, \Delta_g \, , 
\ee
which goes like $(- \Delta_g)^{\gamma +1} \, \Delta_g$ at high energy. 
 The necessary trace 
 to calculate the average return probability is 
obtained from the Fourier transform of (\ref{Kk}), 
\be
K_g(x,x^{\prime}; T) \propto \int {\rm d}^4 k \,  e^{ - k^2 \, \bar{h}(k^2/\Lambda^2) \, T  } \, 
e^{ i k (x - x^{\prime})}.
\ee 
Now we are ready to calculate the average return probability defined in (\ref{trace})
\be
P_g(T) \propto \int {\rm d}^4 k \, e^{- k^2 \, \bar{h}(k^2/\Lambda^2)\, T } . 
\label{PT}
\ee
From the requirement (iii) we know that, at high energy, $h(k) \sim k^{2 \gamma}$ and then
$\bar{h}(k) \sim k^{2 \gamma + 2}$; therefore, we can calculate the integral (\ref{PT}) 
and then the spectral dimension defined in (\ref{dimform}) for small $T$ will be 
\be
P_g(T) \propto 
 \frac{1}{T^{\frac{2}{2 + \gamma}}}  \,\,\,\,\, \Rightarrow \,\,\,\,\,  d_s = \frac{4}{\gamma +2 }.
\ee
The parameter $\gamma\geqslant 3$ implies that the spectral dimension 
is $d_s <1$, manifesting a fractal nature of the spacetime at high energy.

\begin{figure}[ht]
\begin{center}
\includegraphics[width=6cm,angle=0]{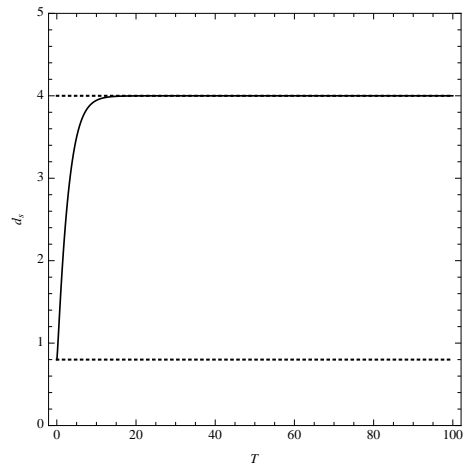}
\caption{\label{SpDim} Plot of the spectral dimension as a function of the fictitious time $T$
for the special case $\gamma=3$ in (\ref{HS}). The lowest value in the picture is $d_s = 4/5$ at high energy,  
but at low energy the spectral dimension flows to $d_s =4$. }
\end{center}
\end{figure}

We can calculate the spectral dimension at all energy scales as a function of the fictitious time $T$
using the explicit form of the entire function $H(k^2/\Lambda^2)$ given in (\ref{HS}).
Integrating numerically (\ref{PT}), we can 
plot directly the spectral dimension achieving the graphical result in 
Fig.\ref{SpDim}\footnote{For the operators 
introduced in the previous section 
$\exp(-\Box/\Lambda^2)^n$ ($n=1,2$),
the propagator scales as 
\be
D(k) \propto \frac{e^{- k^{2 n}/\Lambda^{2 n}}}{k^2},
\label{Dexps}
\ee
and the spectral dimension goes to zero at high energy.
In particular, for $n=1$ the heat-kernel can be calculated analytically, 
\be
K( x ,x^{\prime} ; T)=\frac{e^{- \frac{\left( x-x^{\prime}  \right)^2}{4 (T + 1/\Lambda^2)
 } }}{\left[  4\pi \left( T+1/\Lambda^2 \right) \right]^{2} },
\ee
as it is easy to verify by going back to the propagator (\ref{Dexps}).  
 Now, employing Eq.(\ref{dimform}), we find that the spectral dimension is
\begin{eqnarray}
d_s = \frac{4 \, T}{T + 1/\Lambda^2},
\label{piatto}
\end{eqnarray}
which clearly goes to zero for $T \rightarrow 0$ and approaches $d_s =4$ for $T\rightarrow + \infty$. }.

\section{Black Holes} 
In this section, we want to solve the equation of motion coming from the
renormalized theory in the case of a spherically symmetric spacetime.
Let us start from the 
classical Lagrangian that we rewrite rearranging the parameters in a different way,
\be
&& \hspace{-0.7cm}
\mathcal{L} =
- \sqrt{-g} 
\Big\{  \frac{\beta}{\kappa^2} R  
- \left( \beta_2 - \frac{\alpha_2}{\kappa^2 \Lambda^2}  \right)  (R_{\mu \nu} R^{\mu \nu} - \frac{1}{3} R^2 ) \nonumber \\
&& \hspace{-0.4cm}
+ \left( \beta_0  - \frac{\alpha_0}{6 \kappa^2 \Lambda^2} \right) R^2 
- \alpha \, G^{\mu \nu} \, \frac{\tilde{h} (- \Box_{\Lambda}) - 1 }{\kappa^2 \, \Box} \, R_{\mu \nu} 
 \Big\} ,
\ee
where $\tilde{h}(z) : =\exp H(z)$\footnote{
In general a differential equation with an infinity number of derivative 
has not a well-defined initial value problem and it needs an infinite number 
of initial conditions. It is shown in \cite{barnaba} that in a general framework 
each pole of the propagator contributes 
two initial data to the final solution. This is precisely our case because the only pole 
in the bare propagator is the massless graviton and the theory has a well defined 
Cauchy problem.}. 
From \cite{Barvi}, the equations of motion for the above theory up to square curvature terms are 
\be
G_{\mu \nu} + O(R_{\mu \nu}^2) 
+ O(\nabla_{\mu} \nabla_{\nu} R_{\rho \sigma})= 8 \pi G_N \tilde{h}^{-1}
T_{\mu \nu},
\label{MEE}
\ee
where we omitted the argument of $\tilde{h}( - \Box_{\Lambda})$. 
Since we are going to solve the Einstein equations neglecting curvature square terms, 
then we have to impose the conservation $\nabla^{\mu} (\tilde{h}^{-1}T_{\mu \nu})=0$
in order for the theory to be compatible with the Bianchi identities. For the exact equations of motion the 
Bianchi identities are of course satisfied because of the diffeomorphisms invariance. 
The condition 
$\nabla^{\mu} (\tilde{h}^{-1}T_{\mu \nu})=0$ compensates the truncation in the modified Einstein equations 
(\ref{MEE}). 

\begin{figure}[ht]
\begin{center}
\includegraphics[width=4cm,angle=0]{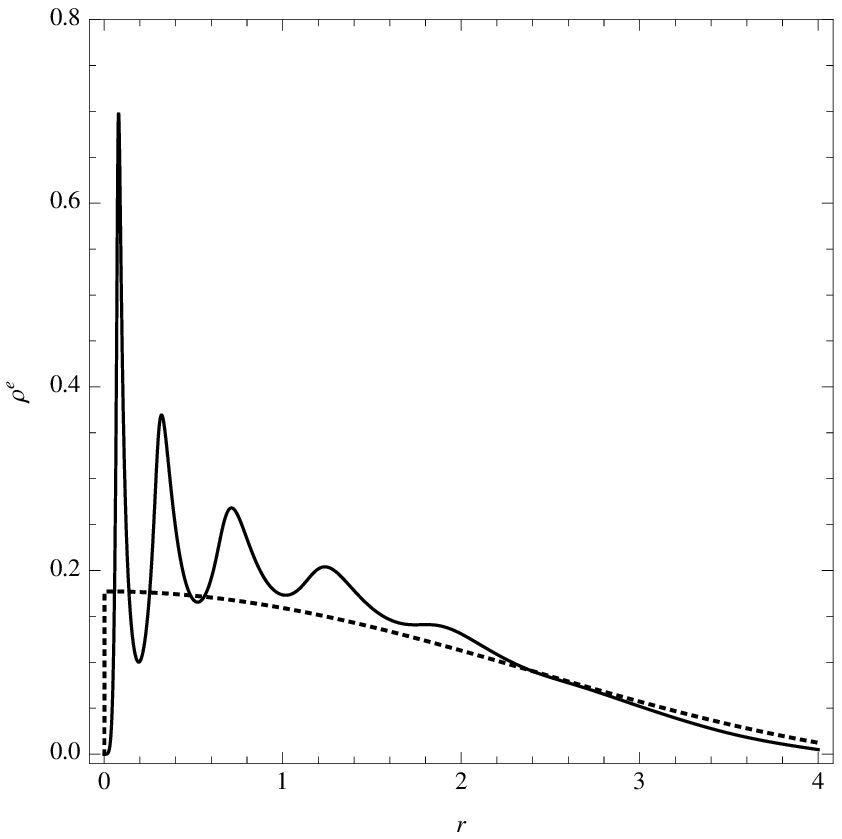}
\includegraphics[width=4cm,angle=0]{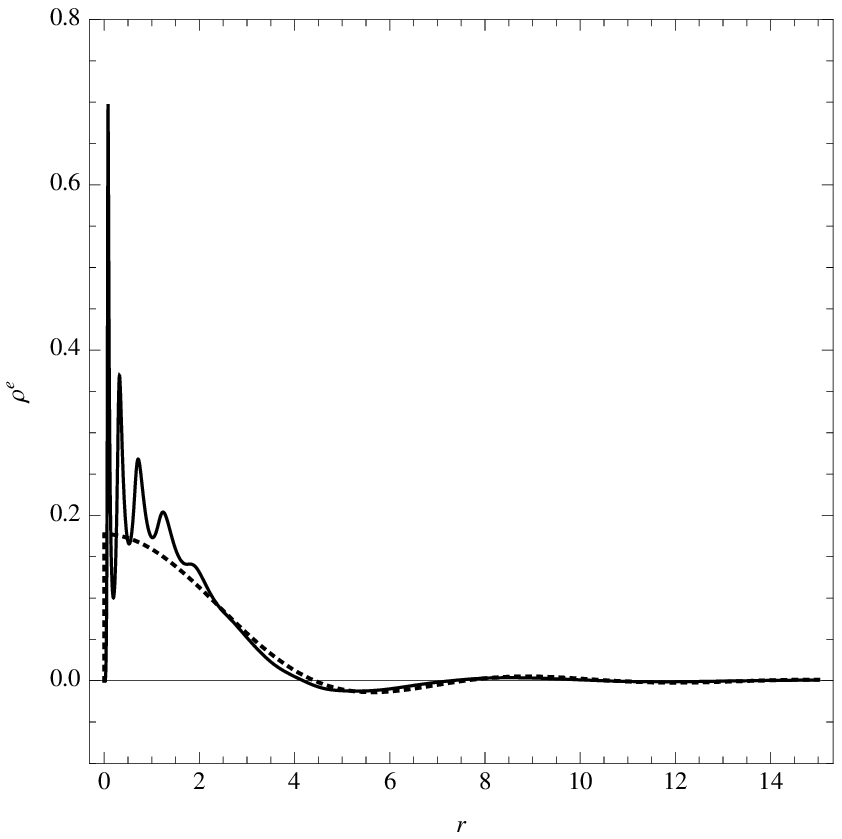}
\caption{\label{plotrhoeff3} Plot of the energy density for $m=10$ in Planck units assuming 
$\Lambda = m_P$. The solid line is a plot of (\ref{rhoeff}) without any approximation, 
the dashed line refers to the 
energy density profile (\ref{rhoeff2}) in the limit $- \Box/\Lambda^2 \ll 1$.}
\end{center}
\end{figure}

Our main purpose is to solve the field equations by assuming
a static source, which means that the four-velocity field $u^\mu$ has only
a non-vanishing time-like component
$u^\mu\equiv ( u^0 , \vec{0} )$ 
$u^0= (- g^{00})^{-1/2}$.
We consider
the component  $T^0\,_0$ of the energy-momentum tensor for a static source of mass $m$
in polar coordinates to be
 $T^0\,_0= - m \delta(r)/4\pi\, r^2 \label{t00}$ \cite{deBene},\cite{SpallucciUnp}\footnote{
Usually, in General Relativity textbooks,  
the Schwarzschild solution is introduced without mentioning the 
presence of a point-like source. 
Once the Einstein equations are solved in the vacuum, the integration constant is determined 
by matching the solution with the Newtonian field outside a spherically symmetric 
mass distribution. Definitely, this is not the most straightforward way to expose 
students 
to one of the most fundamental solutions of the Einstein equations. 
Moreover, the presence of a curvature singularity in the origin, 
where from the very beginning a Þnite mass-energy is squeezed into a zero-volume 
point, is introduced as a shocking, unexpected result. Against this background, 
we show  
that, once quantum delocalization of the source is 
accounted, all these flaws disappear. From this follows that for us there is only one physical 
vacuum solution and this is the Minkowski metric. In other words, the Schwarzschild metric is 
a vacuum solution with the free integration $m$ equal to zero.
}.
The metric of our spacetime is assumed to be given by the usual static, spherically symmetric 
Schwarzschild form
\be
&& ds^2=- F(r)dt^2 + \frac{dr^2}{F(r)}+r^2\Omega^2, \nonumber \\
&& F(r)=1-\frac{2 m(r) }{r}.
\label{metricF}
\ee
The effective energy density and pressures are defined by 
\begin{eqnarray}
 \hspace{-0.5cm} 
\tilde{h}^{-1}T^{\mu}{}_{\nu} 
=\frac{G^{\mu}\hspace{0.001cm}_{ \nu}}{8 \pi G_N} = {\rm Diag}(- \rho^{\rm e}, P_r^{\rm e}, P_{\bot}^{\rm e}, P_{\bot}^{\rm e}).
\end{eqnarray}
For later convenience, we temporarily adopt free-falling Cartesian-like coordinates \cite{SpallucciUnp}, 
we calculate
the effective energy density assuming $p_{\gamma +1}(z) = z^4$ in $H(z)$,
\be
&& \hspace{-0.5cm}  
\rho^{\rm e}( \vec{x}) := - \tilde{h}^{-1}(-\Box(x)_{\Lambda}) T^{0}{}_{0} = 
 m \, \tilde{h}^{-1}(-\Box(x)_{\Lambda})\, \delta(\vec{x})  \nonumber \\
 && \hspace{0.5cm} = m  \int \frac{d^3 k}{(2 \pi)^3} \,   e^{-  H(k^2/\Lambda^2)  } e^{i \vec{k} \cdot \vec{x}  } \nonumber \\
&&   \hspace{0.5cm} = \frac{ 2 m}{(2 \pi)^2 \, r^3} \int_0^{+\infty} e^{- H(p^2/r^2 \Lambda^2)} p \, \sin(p) \, dp,
\label{rhoeff}
\ee
where $r=|\vec{x}|$ is the radial coordinate. Here we introduced the Fourier transform 
for the Dirac delta function and we 
we also introduced a new dimensionless variable in the momentum space, $p = k \, r$, where $k$ 
is the physical momentum. 
The energy density distribution defined in (\ref{rhoeff})
respects spherical symmetry. 
We evaluated numerically the integral in (\ref{rhoeff}) and the resulting
energy density is plotted in Fig.\ref{plotrhoeff3}.
In the low energy limit we can expand $H(z)$ for $z = - \Box/\Lambda^2 \ll 1$ and we can 
 integrate analytically 
(\ref{rhoeff}) 
\be
\rho^{\rm e}( r ) 
 = \frac{ 2 m}{(2 \pi)^2 \, r^3} \int_0^{+\infty} e^{-p^{16}/(2 r^{16} \Lambda^{16})} p \, \sin(p) \, dp. 
\label{rhoeff2}
\ee
The result is really involved and the plot is given in Fig.\ref{plotrhoeff3}; however, 
the Taylor expansion near $r \approx 0$ gives a constant leading order 
\be
\rho^{\rm e}(r) \approx \frac{m \Lambda^3}{32 \, 2^{7/16} \Gamma(11/16) \Gamma(7/8) \Gamma(5/4)} + O(r^2).
\ee

The covariant conservation and the additional condition, $g_{00}=-g_{rr}^{-1}$,
completely specify the form of $\tilde{h}^{-1}T^{\mu}{}_{\nu}$ and  
the Einstein's equations reads 
\begin{eqnarray}
&& \frac{ d m(r)}{dr} = 4 \pi \rho^{\rm e} \, r^2 ,  \nonumber \\
&& \frac{1}{F} \frac{ d F}{dr} = \frac{2 \left( m(r) + 4 \pi  P_r^{\rm e} \, r^3\right)}{r (r-2 m(r))} , \nonumber \\
&& \frac{d P_r^{\rm e}}{d r} = - \frac{1}{2 F}  \frac{ d F}{d r} (\rho^{\rm e} + P_r^{\rm e}) 
+ \frac{2}{r} (P_{\bot}^{\rm e} - P_r^{\rm e} ),
\label{Eineq}
\end{eqnarray}
Because of the complicated energy density profile, it is not easy to integrate the first Einstein equation 
in (\ref{Eineq})
\begin{equation}
m(r) =  4\pi \int_0^r dr' r'^2 \ \rho^{\rm e}(r').
\label{mass}
\end{equation}
 However, the energy density goes to zero at infinity, reproducing the
asymptotic Schwarzschild spacetime with $m(r) \approx m$ (constant). 
On the other hand, it is easy to calculate the energy density profile close to $r \approx 0$
since $H(z) \rightarrow \log z^4$ for $z \rightarrow + \infty$ 
(or $r \rightarrow 0$ in (\ref{rhoeff})). 
In this regime $m(r) \propto m \, \Lambda^8 r^8$ and for a more general monomial 
$p_{\gamma +1}(z) = z^{\gamma +1}$, $m(r) \propto m  (\Lambda \, r)^{2 \gamma +2}$.
The function $F(r)$ in the metric, close to $r\approx 0$, is 
\be
F(r) \approx 1 - c \, m \, \Lambda^{2 \gamma +2} \, r^{2 \gamma +1} ,
\label{core}
\ee
where $c$ is a dimensionless constant. 

We show now that the metric has at least two horizons, an event horizon and a Cauchy horizon.
The metric interpolates two asymptotic flat regions, one at infinity and the
other in $r = 0$, so that we can write the $g_{r r}^{-1} = F$ component in the following way
\be
F(r) = 1 - \frac{2 m f(r)}{r},
\ee
where $f(r) \rightarrow 1$ for $r\rightarrow  \infty$, $f(r) \propto r^{2 \gamma +2}$ for $r\rightarrow 0$
and $f(r)$ does not depend on the mass $m$.
The function $F(r)$ goes to ``$1$" in both limits (for $r \rightarrow + \infty$ and $r \rightarrow 0$)
and, since $m$ is a multiplicative constant, we can always choose the mass $m$ for a fixed value of
the radial coordinate $r$, such that $F(r)$ becomes negative everywhere. From this it follows that the function 
$F(r)$ must change sign at least twice. The second equation in (\ref{Eineq}) is solved by 
$P_r^{\rm e} = - \rho^{\rm e}$ and the third one defines the transversal pressure once known the energy density $\rho^e$.
Given the lapse function $F(r)$ in (\ref{core}), 
we can calculate the Ricci scalar and the Kretschmann invariant
\be
&& \hspace{-1cm}  R = c \, m \, \Lambda^{2 \gamma +2} \, (2 \gamma +2) (2 \gamma +3) \, r^{2 \gamma -1} ,  \\
&& \hspace{-1cm}  R_{\mu\nu\rho\sigma} R^{\mu\nu\rho\sigma}  = \nonumber \\
&& \hspace{-1cm} = 4 \, c^2 \, m^2 \, \Lambda^{4 \gamma +4} \, \left(4 \gamma ^4+4 \gamma ^3+5 \gamma ^2+4 \gamma +2\right) r^{4 \gamma}. \nonumber 
\ee
By evaluating the above curvature tensors at the origin one finds that they are finite 
for $\gamma > 1/2$ and in particular for the minimal super-renormalizable 
theory with $\gamma \geqslant 3$.

\begin{figure}[ht]
\begin{center}
\hspace{-0.1cm}
\includegraphics[width=6cm,angle=0]{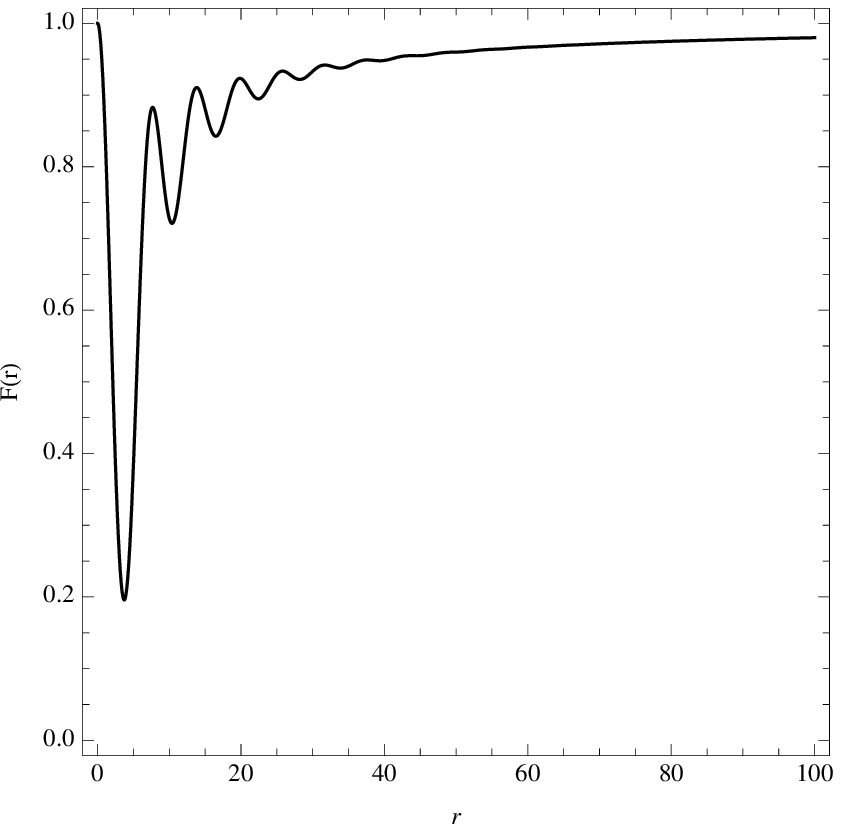}
\includegraphics[width=6cm,angle=0]{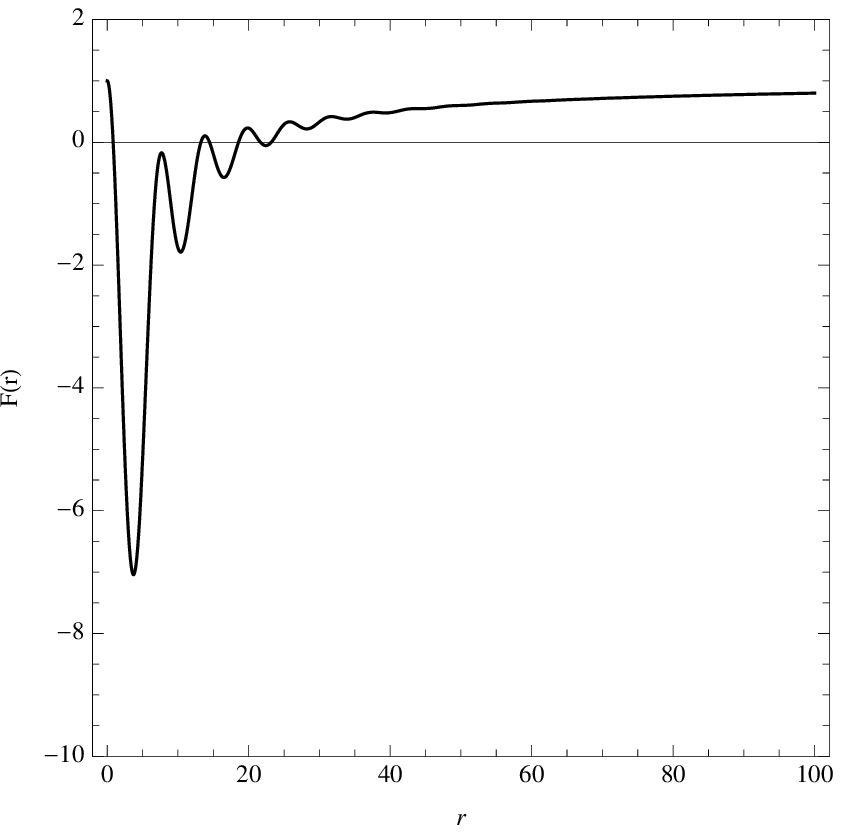}
\caption{\label{Fr1} The two plots show the function $F(r)$, assuming 
the infrared energy density profile (\ref{rhoeff}).
The two plots differ in the value of the ADM mass, which is $m=1$ in the fist plot and $m=10$ 
in the second plot (in Planck units, assuming the fundamental scale $\Lambda$
to be the Planck mass). A crucial property of those black holes is the possibility to have 
``multi-horizon black holes" depending on the mass value. For $m=10$, for example, we have six horizons 
according to the second plot.
}
\end{center}
\end{figure}
\begin{figure}[ht]
\begin{center}
\hspace{-0.1cm}
\includegraphics[width=6cm,angle=0]{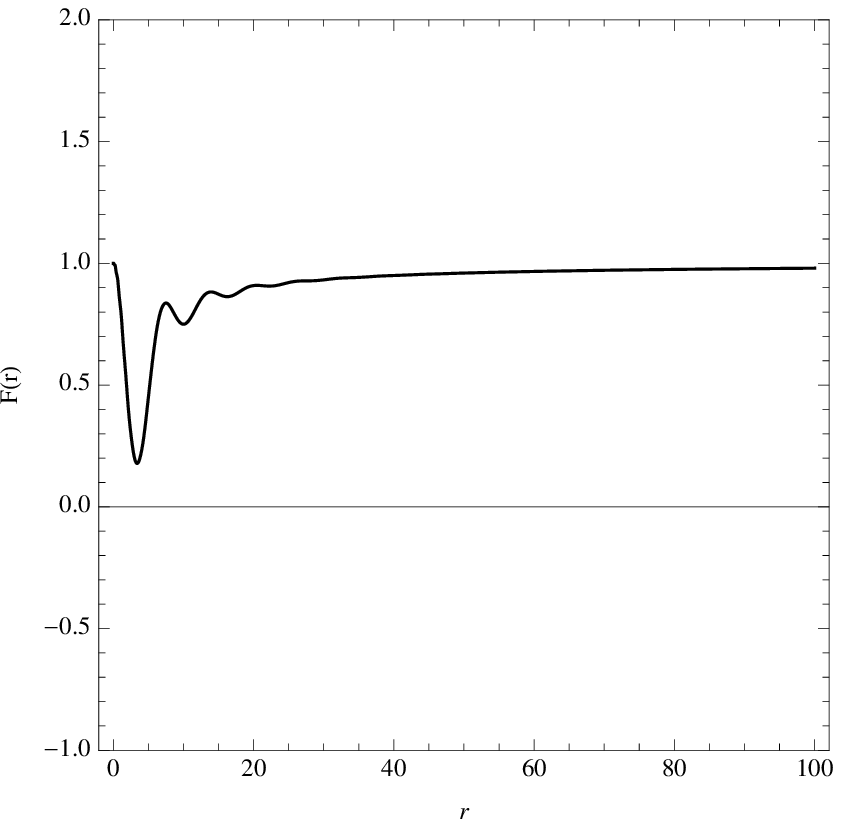}
\includegraphics[width=6cm,angle=0]{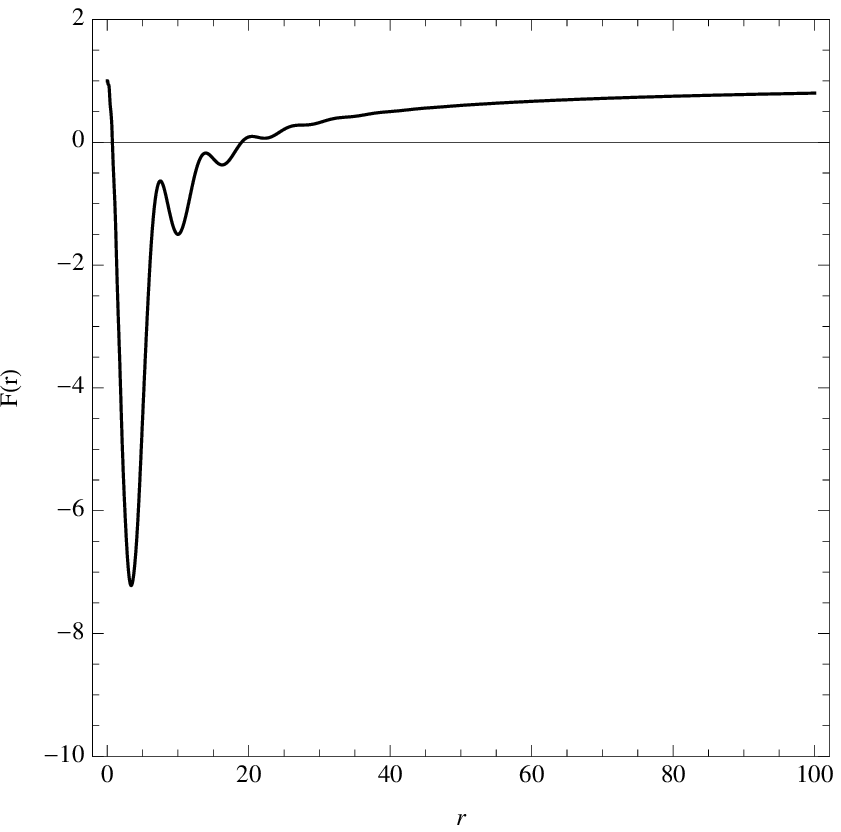}
\caption{\label{Fr2} The first plot shows the function $F(r)$ for the energy profile (\ref{rhoeff2})
and $H(z)$ defined in (\ref{H0}) with the parameter $\gamma =3$.
The ADM mass values are $m=1$ and $m=10$ (in Planck units) for the first and the second plot respectively.
}
\end{center}
\end{figure}
The entire function $h(z)$
is able to tame the curvature singularity of the Schwarzschild solution 
at least for the truncation of the theory here analyzed. We think that 
the higher order corrections to the Einstein equation will not change 
the remarkable feature of the solutions found in this section.

Besides the analysis exposed above, we can integrate numerically the modified Einstein equation of motions 
(\ref{MEE}) for the two energy densities defined respectively in (\ref{rhoeff}) and (\ref{rhoeff2}).
Using the integral form of the mass function (\ref{mass}), we achieve the metric component $F(r)$ 
defined in (\ref{metricF}). 
The numerical results are plotted  
in Fig.\ref{Fr1} and Fig.\ref{Fr2} for different values of the ADM mass $m$. 
The metric function $F(r)$ can intersect zero times, twice or more than twice the horizontal axis 
relative to the value of the ADM mass $m$. This opens the possibility 
to have ``{\em multi-horizon black holes}" as an exact solution of the equation of motions 
(\ref{newgravity}).


\section{Structure of the Interactions} 
We have already shown that the theory is well-defined and power-counting super-renormalizable.  
However, the calculations are not easy beyond the second order in the graviton expansion.
In this section, we give a sketch of how to proceed in the graviton expansion.
%
%
\begin{figure}[ht]
\begin{center}
\hspace{-0.5cm}
\includegraphics[width=4cm,angle=0]{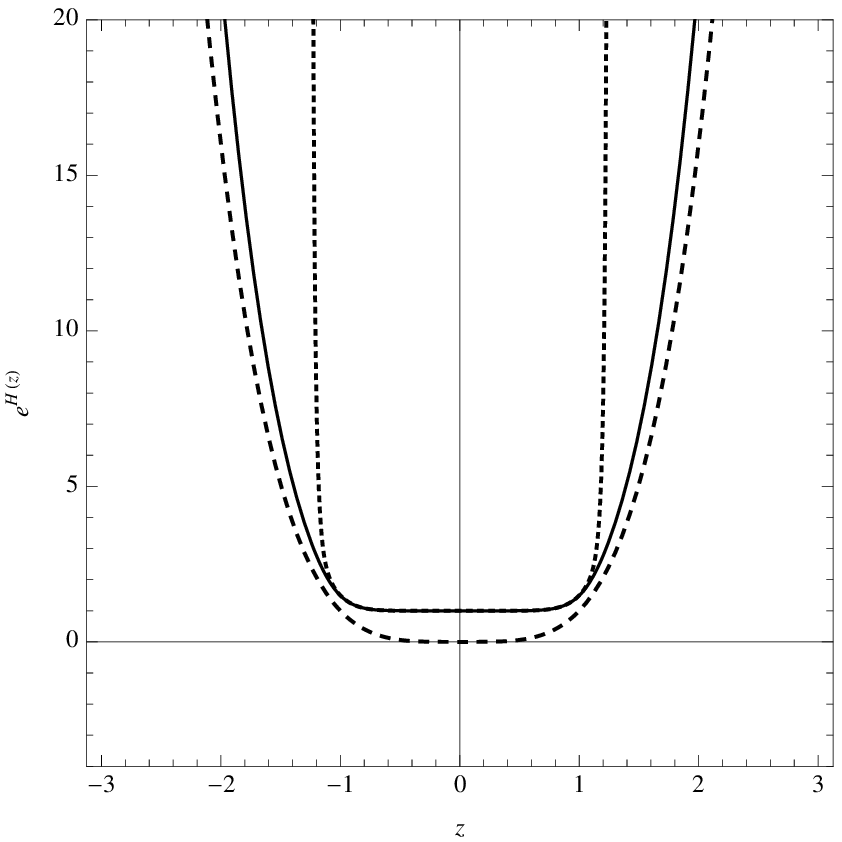}
\includegraphics[width=4cm,angle=0]{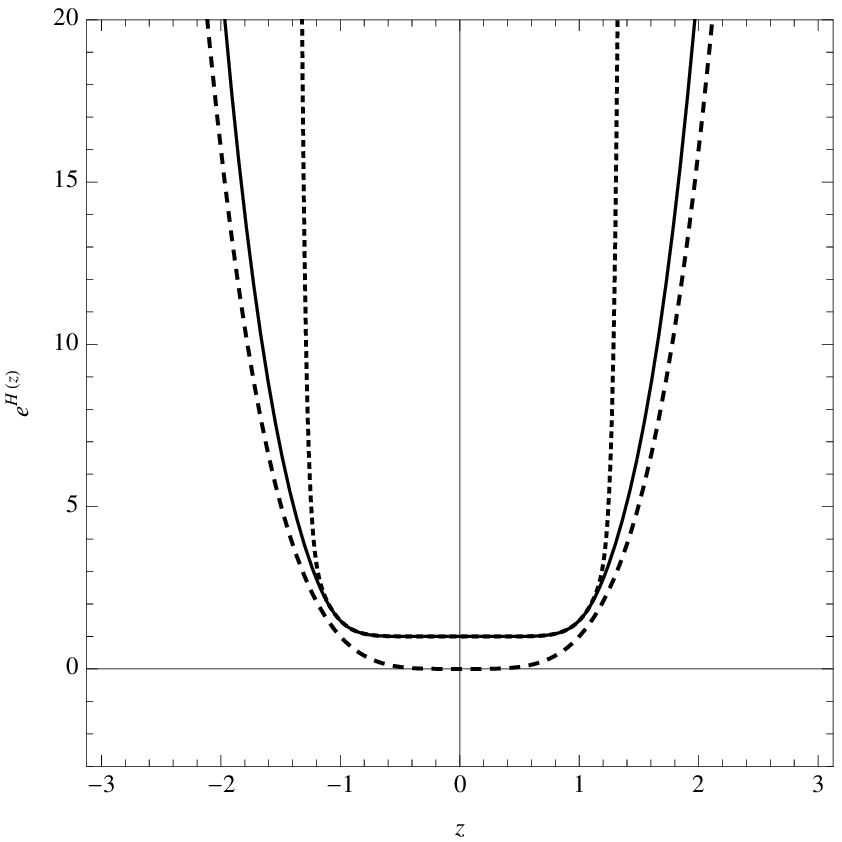}
\caption{\label{ExpHz12} The first plot is the function $\exp H(z)$ for $p_{\gamma +1} = p_4 = z^{4}$.
The solid line represents the exact function, the large and small dashed lines represent the same 
function for large and small value of $z$ respectively: $z^4$ and 
$\exp(z^{24}/36 - z^{16}/8 + z^8/2)$.
In the second plot, the small dashed line represents 
a further simplification of the same function: $\exp H(z) \approx 1 + z^8/2 - z^{24}/72+z^{32}/288$.
}
\label{ExpH}
\end{center}
\end{figure}
The reason for the plots in Fig.\ref{ExpH} is to give an operative definition of $\exp H(z)$.
One possible approximation in the interactions is the following  replacement in the graviton 
expansion for the minimal renormalizable theory with $\gamma  = 3$, 
\be
 e^{H(z)} 
 \approx  \left\{ \begin{array}{ll} z^4 & {\rm for} \,\,\,  z \gtrsim 1.3 \, ,
         \vspace{0.1cm}
         \\   1 + \frac{z^8}{2} - \frac{z^{24}}{72} + \frac{z^{32}}{288}   \,\, &{\rm for} \,\,\, z \lesssim 1.3 \, .
        \end{array} \right. 
        \label{ExpLim}
\ee
At tree level, or loop amplitudes 
at high energy, we can just replace $\exp H(z)$ with $z^4$,  
at low energy with the second expansion defined in (\ref{ExpLim}) for 
$z \lesssim 1.3$ and proceed in the 
calculation, gluing together the results in the two different regimes.

Let us recall again the classical Lagrangian, 
\be
&& \hspace{-0.7cm}
\mathcal{L}= 
- \sqrt{-g} 
\Big\{  \frac{\beta}{\kappa^2} R  
- \left( \beta_2 - \frac{\alpha_2}{\kappa^2 \Lambda^2}  \right)  (R_{\mu \nu} R^{\mu \nu} - \frac{1}{3} R^2 ) \nonumber \\
&& \hspace{-0.8cm}
+ \left( \beta_0  - \frac{\alpha_0}{6 \kappa^2 \Lambda^2} \right) R^2 
- \alpha \, G^{\mu \nu} \, \frac{e^{H( -\Box_{\Lambda})} - 1 }{\kappa^2 \, \Box} \, R_{\mu \nu} 
 \Big\} ,
\ee
where $\exp H(z)$ is defined in (\ref{ExpLim}). 
This Lagrangian interpolates between the Einstein-Hilbert Lagrangian at low energy and 
a high energy theory living in a spacetime of spectral dimension $d_s = 4/(\gamma+2)$.

A first approximation (but also an operative way to proceed) 
is to replace the interaction Lagrangian in the UV with the following truncation
\be
\hspace{-0.5cm}
\mathcal{L}^{\rm int}_{\rm UV} \approx \frac{\alpha}{\kappa^2 \Lambda^2} 
\sqrt{- g} \, G^{\mu \nu} \left( \frac{ - \Box}{\,\,\,\, \Lambda^2} \right)^{\gamma} \, R_{\mu \nu} 
\,\,\, {\rm for} \,\,\, k \gtrsim \Lambda, 
\label{UV}
\ee
($k$ is the energy scale)
and the infra-red Lagrangian with 
\be
&& \hspace{-0.5cm}
\mathcal{L}_{\rm IR}^{\rm int} =
- \sqrt{-g} 
\Big\{  \frac{\beta}{\kappa^2} R  
- \left( \beta_2 - \frac{\alpha_2}{\kappa^2 \Lambda^2}  \right)  (R_{\mu \nu} R^{\mu \nu} - \frac{1}{3} R^2 ) \nonumber \\
&& \hspace{-0.6cm}
+ \left( \beta_0  - \frac{\alpha_0}{6 \kappa^2 \Lambda^2} \right) R^2 
- \frac{\alpha}{2 \kappa^2 \Lambda^{16}} \, G^{\mu \nu} \,  \Box^{2 \gamma +1} \, R_{\mu \nu} 
 \Big\}     
\label{IR}
\ee
for $k \lesssim \Lambda$. 
In (\ref{UV}) and (\ref{IR}) we used the same expansion of (\ref{ExpLim}) 
but for a general value of $\gamma$. 
On the other hand, the propagator is the same in both regimes and is given in (\ref{compPro}).
The three graviton interactions can be obtained 
performing an $h^{\mu \nu}$ power expansion in (\ref{UV}) and (\ref{IR}). 
At tree level, $n$-points functions will be obtain by an interpolation of the amplitude calculated in the
two different regimes $k\lesssim\Lambda$ and $k\gtrsim \Lambda$, using respectively 
the $\mathcal{L}^{\rm int}_{\rm IR}$ and $\mathcal{L}^{\rm int}_{\rm UV}$.
In loop amplitudes, we should integrate the interactions terms coming from $\mathcal{L}^{\rm int}_{\rm IR}$
up to $k \lesssim \Lambda$ and the interactions coming from $\mathcal{L}^{\rm int}_{\rm UV}$
in the range $\Lambda \lesssim k < + \infty$ in the same amplitude.


\section{Conclusions}
In this paper we studied a new perturbative quantum gravity theory with a ``gentle non-local character."
We have shown that it is possible to build a perfectly well-defined quantum gravity theory without extra 
poles in the graviton propagator. 

The properties required for the theory introduced and studied in this paper were the following:
\begin{enumerate}
\renewcommand{\theenumi}{(\roman{enumi})}
\renewcommand{\labelenumi}{\theenumi}
\item the theory should reproduce general relativity in the infra-red limit;
\item black hole solutions of the classical theory have to be singularity free; 
\item the theory should be perturbatively renormalizable or super-renormalizable or finite;
\item the spectral dimension should decrease at short distances;
\item the theory has to be unitary with no other degrees of freedom than the graviton.
\end{enumerate}

All the above properties are satisfied by our quite restrictive class of actions, 
which differ uniquely for the choice of an entire function $H(z)$. 
This class of theories can not be renormalizable or finite but 
such theories turn out to be super-renormalizable, 
since only one loop Feynman diagrams diverge,
implying a renormalization of just three coupling constants. 
The propagator has only one pole in the graviton mass shell.  
The minimal super-renormalizable theory we built has spectral dimension $d_s = 4/5$ 
in the ultra-violet regime and is four dimensional in the infrared limit. 

We have also considered a truncation of the classical theory, showing 
that spherically symmetric black hole solutions 
are singularity-free. As for the solutions in \cite{NS}, we have black holes only if the mass is 
bigger than the Planck mass (if we assume the fundamental scale in the theory to be the Planck mass).
The new black hole solutions are more properly 
``{\em multi-horizon black holes}," showing a very reach spacetime 
structure depending on the value of the ADM mass.

Future work will focus on the following subjects:
\begin{enumerate}[$\bullet$]
\item the cosmological singularity problem at the classical or semiclassical level. Some preliminary 
work has been already done in \cite{Cosmo};
\item the high curvature corrections to the black hole solutions presented in this paper. 
\end{enumerate}
Future work will also take the following directions: 
\begin{enumerate}[$\diamond$]
\item we will consider more in detail  the connection 
between non-locality and the fractality of the spacetime 
(see in particular the recent work by Calcagni in \cite{Carlip}); 
\item we will reconsider super-renormalizable gauge theories, with particular attention to the
grand-unification of the fundamental interactions  
with or without gravity.
It this work, we will try to put together ideas coming from \cite{Tombo} and the more recent paper \cite{GU}.
\end{enumerate}
Possible simplifications of the theory are:
\begin{enumerate}[$\ast$]
\item 
a scalar field  theory with the same non-local structure;
\item a simplification of the metric to the conformal form $g_{\mu \nu} = \Omega^(x) \, \eta_{\mu \nu}$ and 
therefore a quantization of the conformal factor $\Omega(x)$.
\end{enumerate}

\begin{acknowledgments}
\noindent 
We thanks Gianluca Calcagni, Francesco Caravelli, Enore Guadagnini, John Moffat, Alberto Montina, 
Tim Koslowski, Gabor Kunstatter and Pasquale Sodano.
Research at Perimeter Institute is
supported by the Government of Canada through Industry 
Canada and by the Province of Ontario through the
Ministry of Research \& Innovation.
\end{acknowledgments}

\end{document}